\documentclass[twocolumn,showpacs,prb]{revtex4}

\usepackage{graphicx}%
\usepackage{dcolumn}
\usepackage{amsmath}

\makeatletter
\def\btt#1{\texttt{\@backslashchar#1}}%
\DeclareRobustCommand\bblash{\btt{\@backslashchar}}%
\makeatother

\begin{document}

\preprint{HEP/123-qed}

\title[Short Title]{Spectroscopic evidence for a charge-density-wave condensate in a charge-ordered manganite: Observation of collective excitation mode in Pr$_{\text{0.7}}$Ca$_{\text{0.3}}$MnO$_{\text{3}}$ by using THz time-domain spectroscopy}

\author{Noriaki Kida}
\affiliation{Research Center for Superconductor Photonics, Osaka University, 2-1 Yamadaoka, Suita, Osaka 565-0871, Japan}

\affiliation{CREST, Japan Science \& Technology Corporation (JST), 2-1 Yamadaoka, Suita, Osaka 565-0871, Japan}

\author{Masayoshi Tonouchi}

\affiliation{Research Center for Superconductor Photonics, Osaka University, 2-1 Yamadaoka, Suita, Osaka 565-0871, Japan}

\affiliation{CREST, Japan Science \& Technology Corporation (JST), 2-1 Yamadaoka, Suita, Osaka 565-0871, Japan}

\date{\today}

\begin{abstract}\small
THz time-domain spectroscopy was used to directly probe the low-energy (0.5--5 meV) electrodynamics of the charge-ordered manganite Pr$_{0.7}$Ca$_{0.3}$MnO$_3$. We revealed the existence of a finite peak structure around 2--3 meV well below the charge gap $\sim300$ meV. In analogy to the low-energy optical properties of the well-studied low-dimensional materials, we attributed this observed structure to the collective excitation mode arising from the charge-density-wave condensate. This finding provides the importance role of the quasi-one dimensional nature of the charge and orbital ordering in Pr$_{0.7}$Ca$_{0.3}$MnO$_3$.

\end{abstract}

\pacs{75.30.Vn, 71.45.Lr, 72.15.Nj, 73.20.Mf}

\maketitle

\small

\section{Introduction}\label{intro}

\begin{figure}[h]
\includegraphics[trim=185 283 255 -155, clip, keepaspectratio=true, width=6.4cm]{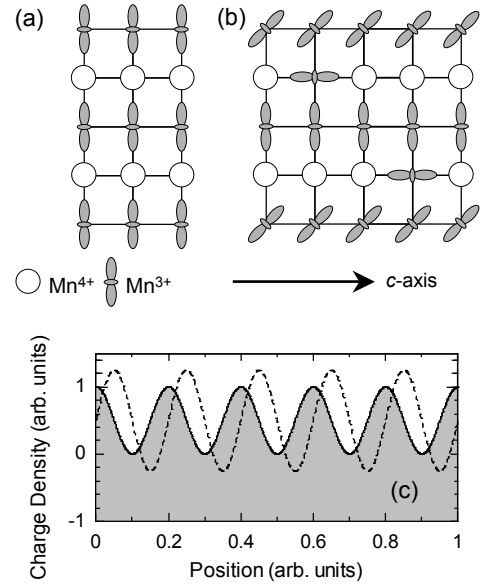}
\caption{\small\baselineskip 7pt Schematic arrangements of the (a) charge exchange (CE)-type charge and orbital ordering pattern observed in half-doped manganites and (b) CE-type structure as proposed in Pr$_{0.625}$Ca$_{0.375}$MnO$_3$ along the $c$-axis by Asaka {\it et al}. (Ref. \cite{TAsaka}). Pr and Ca atoms are omitted to clarify. (c) Schematic illustration of the charge-density-wave formation.}
\label{fig1}
\end{figure}

The tendency toward for the formation of the charge ordering is a common characteristic of the transition metal oxides with a perovskite structure including high-temperature superconducting cuprates and colossal magnetoresistive manganites. \cite{MImada,YTokura} In particular, various kind of the manganites with the doping level $x$ of 1/2 show the charge-exchange (CE)-type charge ordering, \cite{YTokura,ZJirak} in which Mn$^{3+}$ and Mn$^{4+}$ ions by the ratio of 1:1 regularly distribute in the underlying lattice below the charge and orbital ordering temperature $T_{\text{CO/OO}}$ [Fig. \ref{fig1}(a)]. Such a real-space charge ordering is characterized by a single-particle excitation across the charge gap $2\Delta$. Using the light of frequency $\omega$, the single-particle excitation spectrum of the charge-ordered (CO) manganites have been extensively investigated in recent years; \cite{YOkimoto1,YOkimoto2,YOkimoto3,PCalvani,HLLiu,TTonogai} for example, Okimoto {\it et al}. identified $2\Delta\sim300$ meV of Pr$_{0.6}$Ca$_{0.4}$MnO$_3$ by using polarized reflectivity measurements and clarified that the electronic structure is dramatically reconstructed in the order of eV by varying temperature $T$ and magnetic field $H$ (Refs. \onlinecite{YOkimoto2} and \onlinecite{YOkimoto3}). Using transmission measurements, Calvani {\it et al}. found that $2\Delta(T)$ of La$_{1-x}$Ca$_x$MnO$_3$ with $x=0.5$ and 0.67 can be well described by the Bardeen-Cooper-Schrieffer (BCS) relation. \cite{PCalvani}

Another important viewpoint is the collective nature of $e_{\rm g}$ carriers including the orbital degree of freedom, which is thought to have a great impact on extraordinary properties of the manganites, i.e., colossal magnetoresistance (CMR). \cite{YTokura2} The conspicuous examples are recent findings of the ``orbiton'', a new elementary excitation of the orbital-ordered state observed in LaMnO$_3$ \cite{ESaitoh} and of the ``charge stripes'', a new periodic form composed of a pairing of diagonal Mn$^{3+}$ with Mn$^{4+}$ observed in La$_{1-x}$Ca$_x$MnO$_3$ $(x\geq1/2)$. \cite{SMori} Therefore, the understandings of the role of $e_{\rm g}$ carriers with variation of $x$---thus the collective phenomena of $e_{\rm g}$ carriers in connection with the specific kind of the charge ordering depending on $x$---have attracted interests and may be of central topics in CO manganites. More recently, Asaka {\it et al}. proposed the model of the CE-type structure in Pr$_{0.625}$Ca$_{0.375}$MnO$_3$ by using the transmission electron spectroscopy as schematically shown in Fig. \ref{fig1}(b). \cite{TAsaka} It may be viewed as a quasi-one dimensional electronic structure with the reduced dimensionality as compared to the well-known CE-type structure [Fig. \ref{fig1}(a)]: $d_{3x^2-r^2}$ or $d_{3y^2-r^2}$ orbital ordering of Mn$^{3+}$ occurs along $c$-axis. In another angle, $d_{3z^2-r^2}$ orbital of Mn$^{3+}$ was cliped by three Mn$^{4+}$ ions. Such a regular pattern of the distinct charges modifies the uniformity of the charge density, leading to the development of the charge-density-wave (CDW) condensate. \cite{GGruner3} This situation is depicted in Fig. \ref{fig1}(c); the charge density (shared area) periodically varies with the position and also time (dashed line), so the new modulation occurs in the underlying lattice (the period of this modulation is larger than the lattice constant). Such a CDW formation is recently deduced from the one-dimensional Fermi-surface (FS) topology in a layered manganite La$_{1.2}$Sr$_{1.8}$Mn$_2$O$_7$ by Chuang {\it et al}.; \cite{YDChuang} they determined the detailed map of the FS by using the angle-resolved photoemission spectroscopy and concluded that FS is subject to CDW instabilities even in the ferromagnetic metallic state. In the sight of these considerations, the charge ordering is also characterized by a collective excitation well below $2\Delta$. However, no clear evidence for the collective excitation of the CO state was reported so far (see, {\it Note added}).

The collective excitation is associated with the mode depending on both the position and the time; the modes arising from the former and the later are referred to as amplitudons and phasons, respectively. \cite{GGruner1} According to Lee {\it et al}., \cite{PALee} the dispersion relation of the phasons is given by $\omega^2\propto |q|^2$, where $q$ is the wavenumber. By the illumination of light, the collective excitation can make a coupled oscillation, which yields the change of the dielectric constant at $q=0$, being in proportional to $(\omega_{\rm LO}^2-\omega^2)/(\omega_{\rm TO}^2-\omega^2)$, where $\omega_{\rm LO}$ and $\omega_{\rm TO}$ are the longitudinal and transverse optical frequencies, respectively. Therefore, this kind of the coupled mode comes to the surface as a collective excitation mode in the complex optical spectrum. It is also important that the impurity and/or the lattice imperfection has striking influence on the CDW dynamics, because the CDW condensate is pinned by the impurity and therefore the peak position of the collective excitation mode shifts to a finite frequency. \cite{GGruner1} 

In this article we provide the spectroscopic evidence for the CDW condensate in the typical CO manganite Pr$_{0.7}$Ca$_{0.3}$MnO$_3$. The basis of this conclusion is derived from the first observation of the collective excitation mode in the optical conductivity spectrum. As well-known CDW system like a quasi-one dimensional K$_{0.3}$MoO$_3$, the collective excitation mode frequently appears in the millimeter $\omega$ range $\sim$ meV, \cite{GGruner1} where it is difficult to get the probing light by the conventional optical spectroscopy. We overcome this limitation by using THz time-domain spectroscopy (TDS), \cite{MCNuss} which is a powerful tool to unveil the low-energy charge dynamics of the CMR manganites. \cite{NKida1,NKida3} In addition to this advantage, THz-TDS is especially suited to capture the signal of the fluctuation phenomenon as the high $\omega$ THz pulse is sensitive and matches on the typical time scale of the charge fluctuation in correlated electron system, which is instantiated by THz-TDS experiments on underdoped Bi$_2$Sr$_2$CaCu$_2$O$_8$. \cite{JCorson}

{\it Material}---Material we used in this work is the typical CO manganite; Pr$_{0.7}$Ca$_{0.3}$MnO$_3$ with a orthorhombically disordered three dimensional structure. This material exhibits the insulating behavior and undergoes the charge \cite{ZJirak,HYoshizawa} and orbital \cite{ZJirak,MvZimmermann} ordering below $T_{\text{CO/OO}}\sim220$ K and the antiferromagnetic spin ordering below $T_{\text{N}}\sim140$ K. Due to the deviation of $x$ from 1/2, at which the charge ordering is most stable, the extra electrons $(x<1/2)$ occupy at the Mn$^{4+}$ site with maintaining the charge ordering [Fig. \ref{fig1}(b)]. The remarkable characteristics compared to other CO manganites are such an admissibility of the charge ordering and the CMR effect, in which the dramatic variance of the resistance by more than 10 orders of magnitude was attained \cite{YTomioka,YTomioka1} Therefore, various kind of the experimental results are accumulated to date. \cite{YTokura} It should be mentioned that there are variant interpretaions about the low-temperature phase below 110 K: One is due to the coexistence of the ferromagnetic and antiferromagnetic phases proposed by Jir\'{a}k {\it et al}.; \cite{ZJirak} another is the canted antiferromagnetic phase (or the spin-glass phase) reported by Yoshizawa {\it et al}. \cite{HYoshizawa} However, the intimate identification is beyond the scope of this study, but we can say that magnetic properties of our samples are consistent with the results by Deac {\it et al.} \cite{IGDeac}

\begin{figure}[t]
\includegraphics[trim=114 377 188 240, clip, keepaspectratio=true, width=7cm]{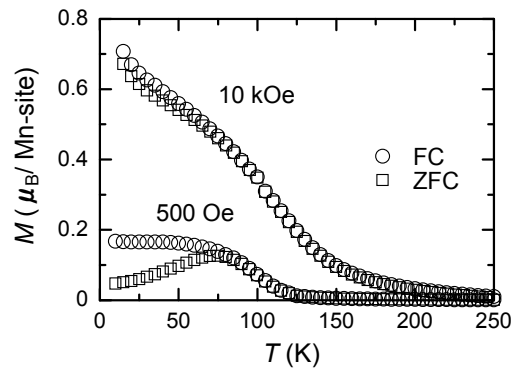}
\caption{\small\baselineskip 7pt Temperature $T$ dependence of the magnetization $M$ in a unit of Bohr magnetron $\mu_{\text{B}}$ at the Mn-site of the sample A. Circles and squares represent $M(T)$ taken from the field cooling (FC) and zero field cooling (ZFC) scans under the magnetic field of 500 Oe or 10 kOe, respectively.}
\label{fig2}
\end{figure}

\begin{table*}[t]
\caption{\small\baselineskip 7pt Magnetic characteristics of the sample A deduced from Fig. \ref{fig2} and the sample B under the magnetic field $H=$ 500 Oe and 10 kOe. The temperature $T_{\rm irr}$ at which the temperature $T$ dependence of the magnetization $M(T)$ in the field cooling (FC) and zero-field cooling (ZFC) scans deviate, $\Delta M=(M_{\rm FC}-M_{\rm ZFC})/M_{\rm FC}$ at 15 K, and $\Delta T=T_{\rm irr}-T_{\rm max}$, where $M_{\rm FC}$ and $M_{\rm ZFC}$ are $M$ in the FC and ZFC scans, respectively, and $T_{\rm max}$ is the maximum temperature in $M_{\rm ZFC}(T)$.
\label{table1}}
\begin{ruledtabular}
\begin{tabular}{ccccccc}
&\multicolumn{3}{c}{Sample A}&\multicolumn{3}{c}{Sample B}\\
\colrule
$H$ (Oe)&$T_{\rm irr}$ (K)&$\Delta M$&$\Delta T$ (K)&$T_{\rm irr}$ (K)&$\Delta M$&$\Delta T$ (K)\\
\colrule
500& 115 & 0.69 & 40 & 115 & 0.82 & 30\\
10,000& 80 & 0.05\footnotemark[1] & 65\footnotemark[1]& 80 & 0.06\footnotemark[1] & 65\footnotemark[1]\\
\end{tabular}
\end{ruledtabular}
\footnotetext[1]{Note, we define $\Delta M$ at 15 K and regard $T_{\rm max}$ as 15 K. In the original report of Deac {\it et al}. (Ref. \onlinecite{IGDeac}), they used these values at 4 K.}
\end{table*}

\section{Methods}

\subsection{Sample preparation and characterization}

Pr$_{0.7}$Ca$_{0.3}$MnO$_3$ thin films on MgO(100) substrates were prepared by a pulsed laser deposition technique using a KrF excimer laser (248 nm in the center wavelength). The thin films were deposited at 800 $^\circ$C under an oxygen pressure $P({\rm O_2})$ of 100 Pa and subsequently cooled down under $P({\rm O_2})$ of 300 Pa. An x-ray diffraction pattern at room temperature indicates that the obtained films were of single phase and were mainly $a (b)$-axis oriented in the cubric setting [we barely detected the (110) intensity peak in the logarithmic intensity scale]. We used two samples (A and B) in the following THz-TDS experiments, both of which are obtained by the same growth condition described above. The film thickness $d$ of the samples A and B are $\sim60$ nm and $\sim100$ nm, respectively. We measured the $d$ dependence of the $a (b)$-axis lattice constant and observed its sudden increase above $\sim100$ nm with the intense (110) intensity peak, implying the partial relaxation of the strained epitaxial growth below $\sim100$ nm.

The obtained films show the typical insulating behavior between 20--300 K according to the DC resistivity $1/\sigma_1(\omega=0)$ measurements with a four-probe method. 

To characterize magnetic properties of our samples, we measured $T$ and $H$ dependences of the magnetization $M$ in a unit of Bohr magnetron $\mu_{\rm B}$ at the Mn-site by a SQUID magnetometer. All data are collected by subtracting the contribution of $M$ of the MgO substrates, which are measured independently. Figure \ref{fig2} shows $T$ dependence of $M$ of the sample A in the field cooling (FC; open circles) $M_{\rm FC}$ and the zero-field cooling (ZFC; open squares) $M_{\rm ZFC}$ under 500 Oe or 10 kOe. With decreasing $T$ under 500 Oe, both $M_{\rm FC}(T)$ and $M_{\rm ZFC}(T)$ curves enhance below 120 K and $M_{\rm ZFC}(T)$ curve starts to deviate $M_{\rm FC}(T)$ curve below the irreversible temperature $T_{\rm irr}\sim80$ K. For further decrease with $T$, $M_{\rm ZFC}(T)$ curve reaches the maximum at temperature $T_{\rm max}\sim75$ K. The difference between $M_{\rm FC}$ and $M_{\rm ZFC}$ at 15 K is estimated to be $\sim0.1$ $\mu_{\rm B}$ at the Mn-site. On the other hand, $M_{\rm FC}(T)$ and $M_{\rm ZFC}(T)$ curves under 10 kOe are hardwired; we observed the continual increase of $M_{\rm ZFC}$ without the broad maximum and that the difference between $M_{\rm FC}$ and $M_{\rm ZFC}$ at 15 K is fairly small. The obtained magnetic properties of the samples A and B are listed in Table \ref{table1} together with $\Delta M=(M_{\rm FC}-M_{\rm ZFC})/M_{\rm FC}$ at 15 K and $\Delta T=T_{\rm irr}-T_{\rm max}$. These quantities and behaviors of $M(T)$ curves both under 500 Oe and 10 kOe are consistent with previous detailed magnetic studies in single crystal and ceramic samples of Pr$_{0.7}$Ca$_{0.3}$MnO$_3$ by Deac {\it et al}. (Ref. \onlinecite{IGDeac}). It is noticed that we employ $\Delta M$ at 15 K and regard $T_{\rm max}$ as 15 K, while Deac {\it et al}. used these values at 4 K. 

\begin{figure}[b]
\includegraphics[trim=232 240 225 -180, clip, keepaspectratio=true, width=6.5cm]{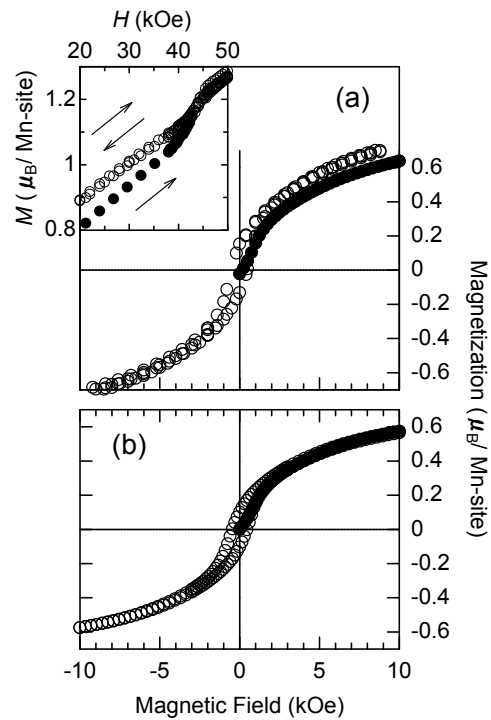}
\caption{\small\baselineskip 7pt Magnetization $M$ in a unit of Bohr magnetron $\mu_{\rm B}$ at the Mn-site as a function of the magnetic field $H$ of the sample A at 30 K. The $H$ cycling is changed between (a) -50 kOe and 50 kOe and (b) -10 kOe and 10 kOe. $M(H)$ curve in the first increasing $H$ and $M(H)$ curve in third increasing and second decreasing $H$ are shown by closed and open circles, respectively. Inset of (a) shows the enlarged view around the metamagnetic transition.}
\label{fig3}
\end{figure}

We show in Figs. \ref{fig3} the $M-H$ curve within $\pm10$ kOe at 30 K after $T$ is decreased under the zero $H$. The $H$ cycling is changed (a) between -50 kOe and 50 kOe and (b) between -10 kOe and 10 kOe. There is a precursor signal of the metamagnetic transition due to the melting of the charge ordering in Fig. \ref{fig3}(a) in contrast to $M-H$ curve shown in Fig. \ref{fig3}(b): \cite{YTomioka} $M$ in the first increasing $H$ (closed circles) does not merge into $M$ in the subsequent third increasing and second decreasing process (open circles); $M$ steeply increases around 45 kOe; Once $H$ exceeds the critical field around 45 kOe [inset of Fig. \ref{fig3}(a)], the magnitude of $M$ in the subsequent increasing and decreasing process is larger than that shown in Fig. \ref{fig3}(b) (see, $M$ at -10 kOe).

We do not directly identify $T_{\text{CO/OO}}$ of our samples by e.g., the transmission electron microscopy. However, as will be shown later, $T$ dependence of the optical conductivity changes around 220 K (see, Fig. \ref{fig6}), which is a nearly same value of $T_{\text{CO/OO}}$ in this material (Ref. \onlinecite{ZJirak}). Moreover, we recently found that $T$ dependence of THz radiation from Pr$_{0.7}$Ca$_{0.3}$MnO$_3$ thin film, which is made by the same growth procedure in this work, dramatically changes at 140 K and 220 K. \cite{NKida2} So we consider that our films have $T_{\text{CO/OO}}$ around 220 K.

\begin{figure*}[t]
\begin{minipage}[t]{.47\textwidth}
\includegraphics[trim=158 379 170 -200, clip, keepaspectratio=true, width=12cm]{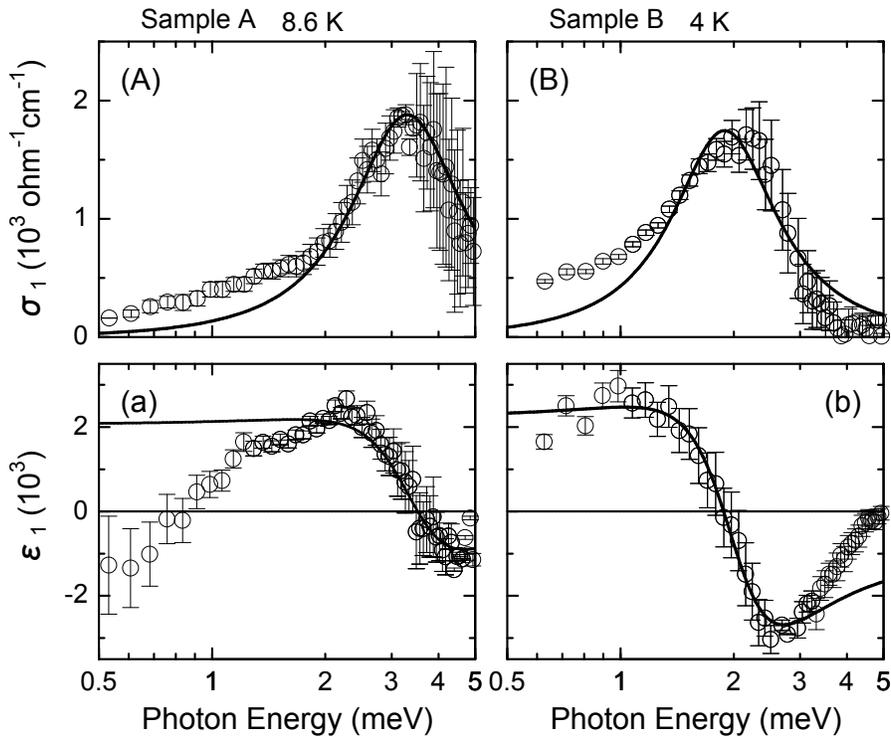}
\end{minipage}
\hfill
\begin{minipage}[b]{.3\textwidth}
\vspace*{-17cm}
\caption{\small\baselineskip 7pt Logarithmic energy plot of the real part of the complex optical conductivity spectrum $\sigma_1(\omega)$ of Pr$_{0.7}$Ca$_{0.3}$MnO$_3$ for (A) sample A and (B) sample B at 8.6 K and 4 K, respectively. The real part of the dielectric spectrum $\epsilon_1(\omega)$ of (a) sample A and (b) sample B are also shown in the lower panel of (A) and (B), respectively. Solid lines in $\sigma_1(\omega)$ and $\epsilon_1(\omega)$ represent fitting results to experimental data (open circles) using Eqs. (\ref{eqn1}) and (\ref{eqn2}) with parameters listed in Table \ref{table2}, respectively.}
\label{fig4}
\end{minipage}
\end{figure*}

\begin{figure*}[tbp]
\includegraphics[trim=13 388 245 7, clip, keepaspectratio=true, width=18cm]{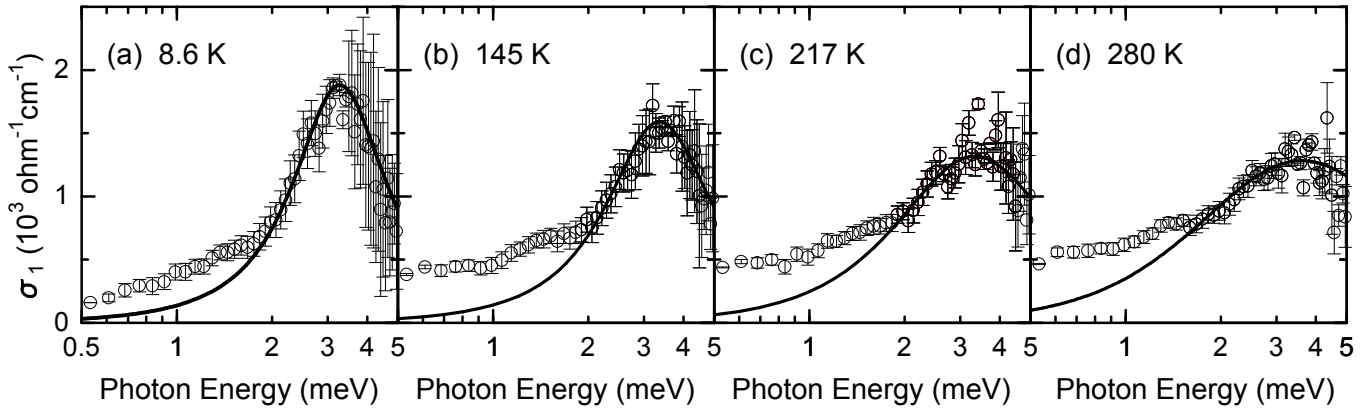}
\caption{\small\baselineskip 7pt Logarithmic energy plot of the real part of the complex optical conductivity spectrum $\sigma_1(\omega)$ of the sample A measured at (a) 8.6 K, (b) 145 K, (c) 217 K, and (d) 280 K. The data are taken from the temperature-warming run. Solid lines represent fitting results to experimental data (open circles) using Eq. (\ref{eqn1}). }
\label{fig5}
\end{figure*}

\begin{figure}[b]
\includegraphics[trim=150 416 135 45, clip, keepaspectratio=true, width=6cm]{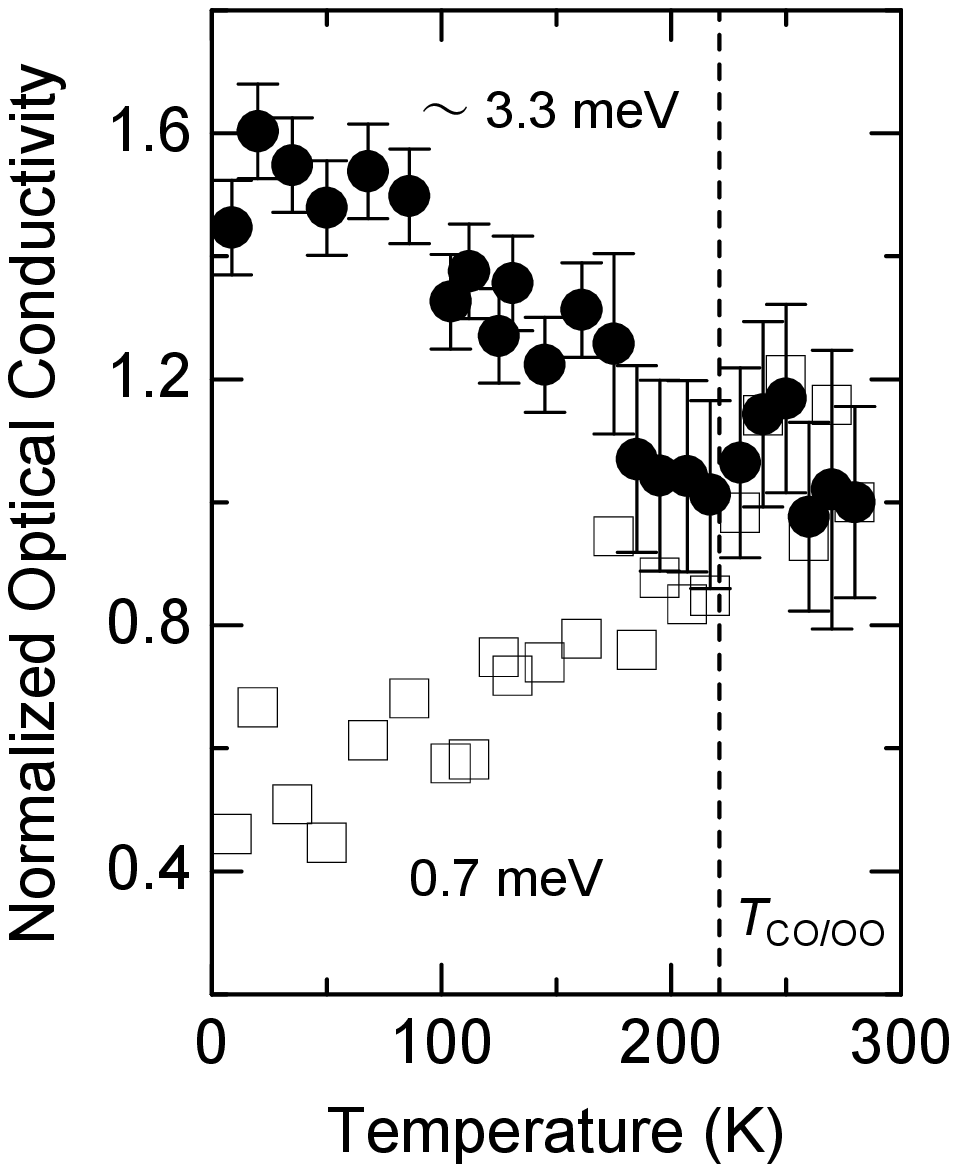}
\caption{\small\baselineskip 7pt Temperature dependence of $\sigma_1(\omega)$ at the peak frequency (closed circles) and at 0.7 meV (open squares) of the sample A. The respective data are normalized by the value at 280 K. The dashed line represents the charge and orbital ordering temperature $T_{\text{CO/OO}}$. }
\label{fig6}
\end{figure}

\subsection{THz time-domain spectroscopy}

We performed THz-TDS experiments utilizing a photoconducting sampling technique in a transmission configuration. THz-TDS can directly provide complex optical spectra without the Kramers-Kronig transformation, which is indispensable relation to estimate complex optical spectra using the conventional optical spectroscopy (reflection or transmission measurements as described in Sec. \ref{intro}). The principle of the THz-TDS can be found in Refs. \onlinecite{MCNuss,NKida1,NKida3}. We briefly describe the experimental setup of our THz-TDS experiments. The light source is a THz pulse from a dipole-type low-temperature grown GaAs (LT-GaAs) photoconducting switch under a voltage bias of 15 V excited by femtosecond optical pulses from the mode-locked Ti:sapphire laser (800 nm in the center wavelength, 150 fs in the pulse width). The THz pulse through the sample is detected by the bow-tie-type LT-GaAs photoconducting switch. In the measured $\omega$ range, optical constants of the MgO substrate is negligible variation with respect to both $\omega$ and $T$. Therefore, we adopt the data of the MgO substrate at room temperature. By performing the Fourier transformation for the transmitted THz pulse, amplitude and phase shift spectra of Pr$_{0.7}$Ca$_{0.3}$MnO$_3$ thin films are obtained. We numerically estimated the complex refractive index spectrum $\tilde{n}(\omega)=n(\omega)-i\kappa(\omega)$ of Pr$_{0.7}$Ca$_{0.3}$MnO$_3$ (detailed procedure, see in Ref. \onlinecite{NKida1}). Then, we transformed to the complex dielectric and complex optical conductivity spectra by using following relations; $\tilde{\epsilon}(\omega)=\epsilon_1(\omega)-i\epsilon_2(\omega)=n(\omega)^2-\kappa(\omega)^2-i2n(\omega)\kappa(\omega)$ and $\tilde{\sigma}(\omega)=\sigma_1(\omega)-i\sigma_2(\omega)=\epsilon_0\omega\Big(\epsilon_2(\omega)-i\big(\epsilon_\infty-\epsilon_1(\omega)\big)\Big)$, respectively, where $\epsilon_\infty$ is the dielectric constant of the bound electrons and $\epsilon_0$ is the permittivity of vacuum.

\section{Results}\label{THz}

Figures \ref{fig4}(A) and \ref{fig4}(B), respectively, show $\sigma_1(\omega)$ with a logarithmic energy scale of Pr$_{0.7}$Ca$_{0.3}$MnO$_3$ for two samples A and B at the low-temperature. We also plot in Figs. \ref{fig4}(a) and \ref{fig4}(b), $\epsilon_1(\omega)$ corresponding to $\sigma_1(\omega)$ of the samples A and B, respectively. The experimental configuration differs from each other, so the light source of THz pulse is different in respective measurements. The open circles denote experimental data. The scattering of the data represented by error bars are large below 0.5 meV and above 4 meV due to a poor sensitivity of the THz light sources and an ambiguous convergence of the transformation procedure. There are important and striking features, which can be clearly seen in the measured $\omega$ range (0.5--5 meV): (i) a finite peak structure appears in $\sigma_1(\omega)$ centered around 2--3 meV below the lowest optical phonon energy $\sim20$ meV (the external optical phonon mode originated from the perovskite structure\cite{YOkimoto3}). Accordingly, the distinguished sharp peaks with a dispersive structure are observed in $\epsilon_1(\omega)$; the peak frequency in $\sigma_1(\omega)$ nearly corresponds to $\epsilon_1(\omega)=0$. (ii) the observed structure has a small spectral weight of the order of 1\% as compared to the single-particle excitation; \cite{YOkimoto2,YOkimoto3,TTonogai} $\sigma_1(\omega)$ at the peak frequency ($\sim 2000$ $\Omega^{-1}$cm$^{-1}$) bears comparison with that at $\sim1$ eV ($\sim 1000$ $\Omega^{-1}$cm$^{-1}$). (iii) its peak position strongly depends on samples, while $\sigma_1(\omega)$ at the peak frequency does not. (iv) the swelling of $\sigma_1(\omega)$ around 1 meV is observed below the finite peak structure. Conformably, $\epsilon_1(\omega)$ continues to exhibit $\omega$ dependence; in particular $\epsilon_1(\omega)$ of the sample A cuts another zero around 0.8 meV.

To reveal the spectral shape of the finite peak structure with $T$, we performed the same experiments in the sample A on warming through $T_{\rm CO/OO}\sim220$ K; $\sigma_1(\omega)$ at selected $T$ [(a) 8.6 K, (b) 145 K, (c) 217 K, and (d) 280 K] are shown in Figs. \ref{fig5}. Fig. \ref{fig5}(a) is same as Fig. \ref{fig4}(A). With increasing $T$ up to $T_{\rm CO/OO}$, $\sigma_1(\omega)$ around the peak frequency decreases and the spectral shape broadens out, while the swelling structure increases [Figs. \ref{fig5}(b) and (c)]. Even above $T_{\rm CO/OO}$, the peak structure is still visible [Fig. \ref{fig5}(d)]. One can notice that the spectral shape around the peak structure at 217 K [Fig. \ref{fig5}(c)] does not considerably change above $T_{\rm CO/OO}$ as can be seen in that at 280 K [Fig. \ref{fig5}(d)]. In addition, the peak frequency doe not depends on $T$.

To clarify the change of $\sigma_1(\omega)$ more clearly with $T$, we plot in Fig. \ref{fig6} the $T$ dependence of the $\sigma_1(\omega)$ at the peak frequency (closed circles) and at 0.7 meV (open squares) of the sample A, which are normalized by the value of the respective $\sigma_1(\omega)$ at 280 K. It can be clearly seen that $\sigma_1(\omega)$ both at the peak frequency and at 0.7 meV above $T_{\text{CO/OO}}$ do not show $T$ dependence within the experimental accuracy [the data of $\sigma_1(\omega)$ around the peak structure is subject to the inaccuracy in the transformation procedure with increasing $T$. Thus, the scattering of the data is large]. On the contrary, with decreasing $T$ below $T_{\text{CO/OO}}$, $\sigma_1(\omega)$ at the peak frequency increases and reaches the maximum value at low-temperature, which is 1.6 times larger than the value at $T_{\text{CO/OO}}$, while $\sigma_1(\omega)$ at 0.7 meV decreases slightly. These results clearly indicate that the observed structure is related to the opening of $2\Delta$.

\section{Assignment and discussion}

Keeping features of the finite peak structure described in Sec. \ref{THz} in mind, let us focus our attention on the assignment of the observed structure.

\begin{figure}[tb]
\includegraphics[trim=130 476 132 105, clip, keepaspectratio=true, width=6.5cm]{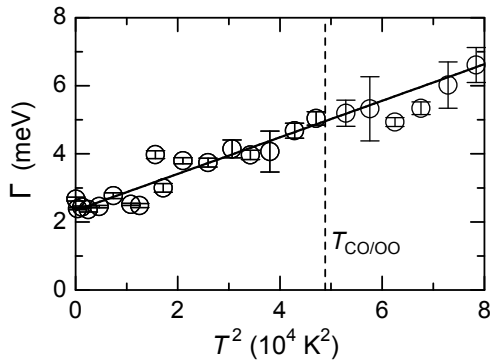}
\caption{\small\baselineskip 7pt Temperature-square $T^2$ dependence of the scattering rate $\Gamma$ of the sample A deduced from Eq. (\ref{eqn1}). The data are taken from the $T$-warming run. The solid line denotes a fitting result to the data (open circles) using Eq. (\ref{eqn3}). The dashed line indicates the charge and orbital ordering temperature $T_{\text{CO/OO}}$.}
\label{fig7}
\end{figure}

At finite $T$, thermal fluctuations in the order of $k_{\text{B}}T$ ($\sim1$ meV at 10 K) as well as the presence of the impurity band play an important role in the measured $\omega$ range, where $k_{\text{B}}$ is the Boltzman's constant; the some hopping conduction of the thermally excited carriers within $2\Delta$ screens in the underlying lattice and subsequently gives rise to a finite peak structure (Ref. \onlinecite{DRomero}). In this viewpoint, $T$ dependence of the scattering rate of the charge carriers in the hopping event follows the exponential $T$ dependence. \cite{SLiu} To see quantitatively how the observed structure changes with $T$, we applied a single Lorentz oscillator expressed as
\begin{equation}
\sigma_1(\omega)=\frac{\sigma_0\omega^2/\tau^2}{(\omega_0^2-\omega^2)^2+\omega^2/\tau^2},
\label{eqn1}
\end{equation}
\begin{equation}
\epsilon_1(\omega)=\epsilon_\infty+\frac{\omega_{\rm p}^2(\omega_0^2-\omega^2)}{(\omega_0^2-\omega^2)^2+\omega^2/\tau^2},
\label{eqn2}
\end{equation}
where $\omega_0$ is the peak frequency, $\tau$ is the relaxation time, $\sigma_0$ is $\sigma_1(\omega)$ at $\omega_0$, and $\omega_{\rm p}$ is the plasma frequency. The solid lines in Figs. \ref{fig4} denote least-squares fits to the respective data. In this procedure, we determined the value of the peak frequency $\omega_0$, where $\epsilon_1(\omega)$ crosses zero and then fixed the value of the optical conductivity $\sigma_0$ at $\omega_0$. Thus only $\tau$ is a fitting parameter in Figs. \ref{fig4}(A) and (B). We also estimate $\epsilon_\infty$ using only above determined values. The obtained quantities of the samples A and B are summarized in Table \ref{table2}. As clearly seen in Figs. \ref{fig4}, we can reproduce $\sigma_1(\omega)$ and $\epsilon_1(\omega)$ by using respective Eqs. (\ref{eqn1}) and (\ref{eqn2}) except for the energy region, where the swelling appears in $\sigma_1(\omega)$. Figure \ref{fig7} displays $T$-square plot of the scattering rate $\Gamma=1/\tau$ of the sample A below 280 K. We found that $\Gamma(T)$ obey the phenomenogical relation given by
\begin{equation}
\Gamma(T)=\Gamma_0+A_2T^2,
\label{eqn3}
\end{equation}
where $A_2$ is the coefficient and $\Gamma_0$ is $\Gamma$ at 0 K. The solid line in Fig. \ref{fig7} is a least-squares fit to the data below $T_{\rm CO/OO}$; Eq. (\ref{eqn3}) holds in good with $\Gamma_0=2.3$ meV and $A_2=0.5$ meV/K$^2$, providing the clear indication that $\Gamma(T)$ cannot be account for by thermal as well as variable range hopping conduction pictures of the carriers in the localized state. \cite{Localization}

\begin{table}[b]
\caption{\small\baselineskip 7pt The obtained quantities deduced from the fitting results to the data shown in Figs. \ref{fig2} by Eqs. (\ref{eqn1}) and (\ref{eqn2}) for different samples A (at 8.6 K) and B (at 4 K); the peak frequency $\omega_0$, the relaxation time $\tau$, the scattering rate $\Gamma=1/\tau$, the optical conductivity $\sigma_0$ at $\omega_0$, and the dielectric constant of the bound electrons $\epsilon_\infty$.
\label{table2}}
\begin{ruledtabular}
\begin{tabular}{cccccc}
Sample&$\omega_0$ (meV)&$\tau$ (ps)&$\Gamma$ (meV)&$\sigma_0$ ($\Omega^{-1}$ cm$^{-1}$)&$\epsilon_\infty$\\
\colrule
A& 3.3 & 0.25 & 2.6 & 1880 & 150\\
B& 1.9 & 0.45 & 1.9 & 1750 & 1000\\
\end{tabular}
\end{ruledtabular}
\end{table}

One may consider the phase separation (PS) picture \cite{MMayr}---the dynamical coexistence of the ferromagnetic metallic and CO insulating states \cite{MUehara}---, which is thought to be a key feature of the CMR phenomena and indeed experimentally discussed in Pr$_{0.7}$Ca$_{0.3}$MnO$_3$. \cite{IGDeac,AAnane,PGRadaelli} Such PS as well as the presence of the disorder commonly seen in thin films using MgO substrates, produces a finite peak structure as a result of the shifting of the Drude peak centered at $\omega=0$. Our samples show the ferromagnetic order at low-temperature; there is a significant hysteresis loop in $M-H$ curve at 30 K with a coercive force of $\sim500$ Oe (Figs. \ref{fig3}), which may imply the presence of the ferromagnetic domain in the CO insulating phase. Therefore, we compare our experimental results to the typical characteristics of the PS in CO manganites. \cite{AAnane,TKimura,RMathieu} Best example in the PS picture is the slow relaxation effect observed in Pr$_{0.67}$Ca$_{0.33}$MnO$_3$ by Anane {\it et al}. (Ref. \onlinecite{AAnane}) and Cr-doped manganites by Kimura {\it et al}., \cite{TKimura} in which the ferromagnetic fraction can be artificially controlled by $H$-annealing process and also the Cr (impurity)-doping at the Mn-site. In these materials, the physical quantities [i.e., $\sigma_1(\omega=0)$ and $M$] strongly depend on the history of the external perturbations. For example, Anane {\it et al}. measured the time dependence of $\sigma_1(\omega=0)$ in the ferromagnetic phase of Pr$_{0.67}$Ca$_{0.33}$MnO$_3$ after $H$ of 100 kOe is applied at low-temperature. \cite{AAnane} They found that the CO insulating phase is restored in the time range of $10^2$--$10^5$ second (the slow relaxation effect). Very recently, Mathieu {\it et al}. reported that $M_{\rm FC}$ under 20 Oe in Nd$_{0.5}$Ca$_{0.5}$MnO$_3$ exhibits the different relaxation effect in the typical characteristic time of $\sim10^2$ second by changing only the cooling rate. \cite{RMathieu} Moreover, the difference of $M_{\rm FC}$ in the history of the cooling rate increases with time. Based on these studies, we have done following THz-TDS experiments: We rapidly cooled the another sample used in Sec. \ref{THz} (the growth condition is same as samples A and B) from room temperature to 4 K with a cooling rate of $\sim8$ K per minute and measured the time dependence of $\sigma_1(\omega)$ while keeping $T=4$ K constant. We observed no slow relaxation effect in the measured $\omega$ range; the spectral shape of the peak structure and the peak frequency $\sim2$ meV exhibit no variation per $\sim10^4$ second, which is a contrarious evidence for the PS picture.

As described in Sec. \ref{intro}, we infer the possibility of the CDW condensate due to the quasi-one dimensional charge and orbital ordering in Pr$_{0.7}$Ca$_{0.3}$MnO$_3$. We discuss here that the observed peak structure arises from the CDW condensate as compared to the low-energy optical properties of the well-known CDW system. The typical characteristics of the collective excitation mode observed in low-dimensional CDW materials are as follows: \cite{GGruner1,TWKim1,GMihaly,TWKim2,LDegiorgi1,LDegiorgi2} (i) As mentioned earlier, the pinning frequency is in general of the order of meV. \cite{GGruner1,TWKim1,GMihaly,TWKim2,LDegiorgi1,LDegiorgi2} (ii) The spectral weight of the collective excitation mode is typically 2 orders of magnitude smaller than that of the single-particle excitation. \cite{GGruner1,TWKim1,GMihaly,TWKim2,LDegiorgi1,LDegiorgi2} This is due to the relatively large $2\Delta\sim100$ meV, which is in strong contrast to the small spin gap $\sim10$ meV observed in the spin-density-wave (SDW) system. In the SDW system, the spectral weight of the collective excitation mode should be comparable to that of the single-particle excitation due to the electron-electron interaction.\cite{GGruner2} (iii) Kim {\it et al}. found that the pinning frequency in (Ta$_{1-z}$Nb$_z$Se$_4$)$_2$I alloys linearly shifts to higher $\omega$ and $\sigma_1(\omega)$ at the pinning frequency does not change with increasing the impurity concentration $z$. \cite{TWKim2} Same tendencies are also observed in K$_{0.3}$Mo$_{1-z}$W$_z$O$_3$ alloys. \cite{LDegiorgi2} (iv) Mih\'{a}ly {\it et al}. identified the swelling structure in the low-energy side of the collective excitation mode in K$_{0.3}$MoO$_3$ as the internal deformation on the basis of the generalized Debye analysis. \cite{GMihaly} Our present findings (i)-(iv) described in Sec. \ref{THz} and various works (i)-(iv) described above in the CDW system, both of which have apparent similar properties and have no contrariety. Moreover, $T^2$ dependence of $\Gamma$ (Fig. \ref{fig7}) has been also found in the CDW system \cite{GGruner1,TWKim1} and Eqs. (\ref{eqn1}) and (\ref{eqn2}) are same expressions derived from the equation of motion for the CDW dynamics. \cite{GGruner1} Therefore, we arrive at the conclusion that the observed structure in Pr$_{0.7}$Ca$_{0.3}$MnO$_3$ is assigned to the collective excitation mode arising from the CDW condensate. To the best of our knowledge, this is a first spectroscopic evidence for the CDW condensate in CO manganites.

The description of the CDW condensate in Pr$_{0.7}$Ca$_{0.3}$MnO$_3$ is consistent and may explain following previous studies. The CDW easily couples with the lattice and affects phonon modes. Therefore, new infrared modes are expected to be active below $T_{\text{CO/OO}}$. Okimoto {\it et al}. found that in addition to phonon modes due to the orthorhombic distortion, many new modes emerge as sharp excitations in Pr$_{0.6}$Ca$_{0.4}$MnO$_3$ at 10 K. \cite{YOkimoto3} Asamitsu {\it et al}. have measured current-voltage characteristics of Pr$_{0.7}$Ca$_{0.3}$MnO$_3$ and found the beginning of the nonlinear conduction above the threshold electric field, \cite{AAsamitsu} which reminds us of the sudden motion of the CDW (CDW depinning) as observed in CDW materials. \cite{GGruner1} 

We should note the previous work about the impurity effect on $\sigma_1(\omega)$ in a copper oxide superconductor. Basov {\it et al}. reported that the Zn (impurity)-doping of the Cu-site in YBa$_2$Cu$_4$O$_8$ distracts the superconductivity and induced the finite peak structure around 10 meV (Ref. \onlinecite{DNBasov}). They concluded that this structure arises from the localization of the charge carriers produced by the randam distribution of the Zn on the Cu-site (disorder) on the basis that the integrated area of $\sigma_1(\omega)$ up to 1 eV conserves as far as 3.5\% Zn-doping. However, same authors also pointed out that if the Drude-like response in YBa$_2$Cu$_4$O$_8$ appears as a result of the collective excitation which is prospective by theory, the peak structure in YBa$_2$(Cu$_{1-z}$Zn$_z$)$_4$O$_8$ can be regard as the collective excitation mode as in both cases of the CDW and SDW. \cite{DNBasov}

It is well known that the collective excitation mode originates in two different states; one is a ``pinned" state due to the pinning of the CDW condensate by the impurity and/or the lattice imperfection and another is the ``bound" state, which is created by the coupling of the pinned state with the optical phonon or the impurity near the pinned state. \cite{LDegiorgi1,LDegiorgi2} Despite the fact that the swelling of $\sigma_1(\omega)$ below the low-energy side of the finite structure, which is usually ascribed to a internal deformation of the pinned collective excitation mode, \cite{GMihaly} can be seen in Figs. \ref{fig4}, we cannot clearly claim whether the observed mode is assigned to the pinned or the bound collective excitation mode of the CDW. One pregnant result is that $\epsilon_1(\omega)$ below the peak structure crosses zero around 8 meV and reaches the minus value with decreasing $\omega$ [Fig. \ref{fig4}(a)]. This implies the presence of another $\omega_{\rm p}$ and of the another peak structure below the measured $\omega$ range. Further experiments on $\tilde{\sigma}(\omega)$ in the GHz $\omega$ range are necessary to perform the detailed discussion and are now in progress using the cavity perturbation technique.

As shown in Fig. \ref{fig5}(d), the collective excitation mode can be seen even above $T_{\text{CO/OO}}$, while its spectral shape is blurred due to the large proportion of the background contribution above $T_{\text{CO/OO}}$. $\Gamma$ continues to follow $T^2$ dependence above $T_{\rm CO/OO}$ as shown in Fig. \ref{fig7} [the fitting procedure by Eq. (\ref{eqn3}) was performed using the data below $T_{\rm CO/OO}$]. This excludes the thermal fluctuation as the origin of this broadening, leading us to conclude that the dynamical short-range CO (charge fluctuation) is still subsistent above $T_{\text{CO/OO}}$ instead of the long-range CO below $T_{\text{CO/OO}}$. Such a short-range CO has been already reported by other experiments: Radaeli {\it et al}. and Shimomura {\it et al}. have revealed that the short-range CO of Pr$_{1-x}$Ca$_x$MnO$_3$ with $x=0.3$, \cite{PGRadaelli} 0.35, 0.4, and 0.5, \cite{SShimomura} exists even above $T_{\text{CO/OO}}$ by neutron and x-ray scattering experiments, respectively. As in the case of the CDW system such as K$_{0.3}$Mo$_{1-z}$W$_z$O$_3$ and (TaSe$_4$)$_2$I, the collective excitation modes are also observed above the CDW transition temperature $T_{\text{CDW}}$ due to the presence of the fluctuating CDW segments, which are systematically investigated by Schwartz {\it et al}. \cite{ASchwartz} For example, the collective excitation mode of K$_{0.3}$MoO$_3$ is visible even at 300 K ($T_{\text{CDW}}\sim183$ K).

Finally, we comment on THz radiation from Pr$_{0.7}$Ca$_{0.3}$MnO$_3$ thin films excited by the femtosecond optical pulses, which we found recently; \cite{NKida2} the radiated spectrum decreases rapidly in intensity with increasing $\omega$ and seems to show the depletion around 2.4 meV. As compared to Figs. \ref{fig4}, such a THz response is due to the absorption of the collective excitation mode during the propagation of the generated THz pulse inside the material.

\section{Summary and outlook}

Summarizing, by using THz-TDS, for the first time, we reported the presence of the finite peak structure around 2--3 meV in Pr$_{0.7}$Ca$_{0.3}$MnO$_3$ and assigned it to the collective excitation mode arising from the CDW condensate. The measurements on the polarization dependence of $\sigma_1(\omega)$ with the grazing incidence of light using the single-domain single crystal will provide the direct evidence and more detailed information for the collective excitation mode. In especially, the measurements of $\sigma_1(\omega)$ in the GHz $\omega$ range are needed to perform further quantitative discussions. Further interests are to study the melting process under external perturbations and $x$ dependence of the collective excitation mode; for example, $2\Delta$ decreases linearly with $x$ from 0.3 to 0.5, \cite{TTonogai} whereas $T_{\text{CO/OO}}$ gradually increases and behaves the less-$x$ dependent. \cite{YTomioka1} This indicates the breaking of the BCS relation given by $2\Delta(x)\propto T_{\text{CO/OO}}(x)$. So, it is indispensable to clarify how the collective excitation mode manifesting well below $2\Delta$ changes with $x$ using THz-TDS.

{\it Note added.}---After the submission of this article, we became an aware of some reports related in the present study. Kitano {\it et al}. [Europhys. Lett. {\bf 56}, 434 (2001)] and Gorshunov {\it et al}. [cond-mat/0201413 (unpublished)] found peak structures around 0.1--1 meV in spin-ladder materials Sr$_{14-x}$Ca$_x$Cu$_{24}$O$_{41}$ by means of optical spectroscopy and assigned them to the collective excitation modes due to the CDW formation. About manganites, Campbell {\it et al}. [Phys. Rev. B {\bf 65}, 014427 (2001)] interpreted that CDW fluctuations inhere in a layered ferromagnetic manganite La$_{1.2}$Sr$_{1.8}$Mn$_2$O$_7$ based on their diffuse x-ray scattering data. Recently, Nagai {\it et al}. [Phys. Rev. B {\bf 65}, 060405(R) (2002)] reported  that the electron microscopic data of a layered CO manganite Nd$_{1-x}$Sr$_{1+x}$MnO$_4$ are consistent with assumptive images of the CDW formation. Very recently, the finite peak structure around 4 meV was also found in 1/8 hole-doped La$_{1.275}$Nd$_{0.6}$Sr$_{0.125}$CuO$_4$ by Dumm {\it et al}. [Phys. Rev. Lett. {\bf 88}, 147003 (2002)] as in the case of YBa$_2$(Cu$_{1-z}$Zn$_z$)$_4$O$_8$. They also ascribe this structure to the localization of the charge carriers in the static stripe phase due to the lack of $2\Delta$ in this material. However, Fujita {\it et al}. reported the evidence for CDW and SDW formations in 1/8 hole-doped La$_{1.875}$Ba$_{0.125-x}$Sr$_x$CuO$_4$ by elastic neutron scattering experiments [Phys. Rev. Lett. {\bf 88}, 167008 (2002)].

\section{Acknowledgements}

We would like to thank in particular Y. Okimoto and also K. Miyano, Y. Ogimoto, and Y. Tokura for giving fruitful comments and discussions. We are also grateful to T. Kawai for giving an opportunity of using SQUID apparatus, T. Kanki, I. Kawayama, K. Takahashi, and H. Tanaka for their help to SQUID measurements, and M. Misra for reading the manuscript.

\end{document}